\documentclass[superscriptaddress,
reprint,
amsmath,amssymb,
aps,
pra
]{revtex4-1}

\usepackage{graphicx}% Include figure files
\usepackage{dcolumn}% Align table columns on decimal point
\usepackage{bm}% bold math
\usepackage{hyperref}% add hypertext capabilities
\usepackage[english]{babel}
\usepackage{color}
\usepackage{epstopdf}

\begin{document}

\title{\color{blue} Thermodynamics and dynamics of two-dimensional systems with dipole-like repulsive interactions}

\author{Sergey A. Khrapak}
\email{Sergey.Khrapak@dlr.de}
\affiliation{Aix Marseille University, CNRS, PIIM, 13397 Marseille,  France} \affiliation{Institut f\"ur Materialphysik im Weltraum, Deutsches Zentrum f\"ur Luft- und Raumfahrt (DLR), 82234 We{\ss}ling, Germany}
\affiliation{Joint Institute for High Temperatures, Russian Academy of Sciences, 125412 Moscow, Russia}

\author{Nikita P. Kryuchkov}
\affiliation{Bauman Moscow State Technical University, 2nd Baumanskaya street~5, 105005 Moscow, Russia}

\author{Stanislav O. Yurchenko}
\email{st.yurchenko@mail.ru}
\affiliation {Bauman Moscow State Technical University, 2nd Baumanskaya street~5, 105005 Moscow, Russia}

\date{\today}

\begin{abstract}
Thermodynamics and dynamics of a classical two-dimensional system with dipole-like isotropic repulsive interactions are studied systematically using extensive molecular dynamics (MD) simulations supplemented by appropriate theoretical approximations. This interaction potential, which decays as an inverse cube of the interparticle distance, belongs to the class of very soft long-ranged interactions. As a result, the investigated system exhibits certain universal properties that are also shared by other related soft-interacting particle systems (like, for instance, the one-component plasma and weakly screened Coulomb systems). These universalities are explored in this article to construct a simple and reliable description of the system thermodynamics. In particular, Helmholtz free energies of the fluid and solid phases are derived, from which the location of the fluid-solid coexistence is determined. The quasi-crystalline approximation is applied to the description of collective modes in dipole fluids. Its simplification, previously validated on strongly coupled plasma fluids, is used to derive explicit analytic dispersion relations for the longitudinal and transverse wave modes, which compare satisfactory with those obtained from direct MD simulations in the long-wavelength regime. Sound velocities of the dipole fluids and solids are derived and analyzed. 
\end{abstract}

\maketitle
%\tableofcontents

\section{Introduction}

Two- and quasi-two-dimensional (2D) interacting particle systems attract great scientific interest, since they play an important role in a broad range of phenomena operating at fluid and solid surfaces and various interfaces~\cite{Kosterlitz1978,KosterlitzRMP2017}. Several relevant examples include atomic monolayers and thin films on a substrate, two-dimensional electron gas on the surface of liquid helium, vortices in thin-film semiconductors, metallic and magnetic layer compounds, smectic liquid crystals, colloidal particles at flat interfaces, and complex (dusty) plasma systems in ground-based conditions.

In the context of colloidal systems, apart from technological applications of colloidally-stabilized emulsions~\cite{CleggJPCM2008, StoccoSoftMatter2009, GarbinPRL2015} and bubbles \cite{PoulichetPNAS2015, YurchenkoLangmuir2016}  for synthesis of novel optical materials \cite{MeerSoftMatter2016, ElsnerJCP2009, GrayJPCM2015, GarbinPRL2015}, chemical sensors, and catalysis~\cite{WonmokLangmuir5262}, many biologically important processes occur at interfaces. Importantly, there is
a way to control these processes by attaching colloidal particles to soft matter interfaces.

When colloidal particles are trapped in oil/water or gas/water interfaces,
electrical dipoles are usually associated with each interfacial particle~\cite{PieranskiPRL1980}. As a result, the interaction between colloidal particles is similar to that between vertically oriented dipoles~\cite{Goulding1998,OettelPRE2007,ParkSoftMatter2011,KelleherPRE2015,KelleherPRE2017} and can be in the first approximation described by a pairwise repulsive inverse-power law (IPL) potential decaying as $\sim 1/r^{3}$ with the interparticle separation $r$. Direct experimental measurements of colloidal interaction potential in such systems by the laser tweezers method \cite{ParkSoftMatter2011, AveyardPRL2002, KelleherPRE2015} or using other approaches \cite{BonalesLangmuir2011} generally confirm this assumption, although it is also clear that actual interactions can be very complicated, particularly in the regime where the separation is comparable to the particle size~\cite{GirottoJPCB2016}. A similar shape of the interaction potential is observed in two-dimensional colloidal systems of paramagnetic particles exposed to an external magnetic field~\cite{ZahnPRL1999, ZahnPRL2000, GasserCPC2010}. 2D colloidal suspensions in external electric fields represent another important class of dipole-like interacting systems.
An external electric field polarizes the particles and ion clouds in the solvent around them, inducing a (tunable) dipole-dipole interaction between the particles. Depending on the orientation of the external electric field with respect to the plane of particle confinement,
the dipolar interaction potential can be either attractive\cite{JuarezJCP2009,CarstensenPRE2015,YakovlevSciRep2017,OvcharovJPCC2017}  or repulsive \cite{DutcherPRL2013,TanakaPRE2014,MaSM2014,WoehlPRX2015,
SainiLangmuir2016,LotitoACIS2017,GongLangmuir2017}.

In the context of plasma physics, it has been long known that the effective potential of a point test charge immersed in a flowing collisionless plasma is not screened exponentially, but falls off as $\propto 1/r^{3}$, at large
distances from the test charge~\cite{Montgomery1968,Cooper1969}.
This can be relevant to complex (dusty) plasmas, a collection of small solid particles in the neutralizing plasma medium. In a typical laboratory dusty plasma experiment the highly negatively charged identical micron-size particles form a horizontal (quasi 2D) layer above the bottom negatively biased electrode of a radio-frequency gas discharge, where the electric force directed upwards is able to balance the gravity force acting on the particles. A strong electric field required to balance the gravity produces significant ion flow, which makes electric potential distribution around the particles highly anisotropic. Although the actual interactions between the particles in these conditions are quite sophisticated and are governed by a competition between screening and plasma-wake mediated effects~\cite{VladimirovPRE1995,ivlev.book,FortovUFN,FortovPR2005,
ChaudhuriIEEE2010,HutchinsonPoP2011,LudwigNJP2012}, there is a certain parameter regime, where the IPL scaling $\propto 1/r^{3}$ is relevant~\cite{KompaneetsPoP2009,KompaneetsPRL2016,KompaneetsPRE2016} (see, in particular, Fig. 3 in Ref.~\cite{KompaneetsPRE2016}).

Thus, dipole-like interactions occur in various two-dimensional physical systems such as ions and colloidal particles trapped at various interfaces, colloidal particles in external electric fields, paramagnetic particles exposed to external magnetic fields, electrical charges placed in a flowing collisionless plasma, etc.  Not surprisingly, structural and dynamical properties, thermodynamics, phase transitions, collective motion and related phenomena in classical systems with $\propto 1/r^{3}$ repulsion have been extensively studied (see, for instance, Refs.~\cite{GrunbergPRL2004,KeimPRL2004,LinPRE2006,LowenPRE1996,vanTeeffelenEPL2006,
vanTeeffelenJPCM2008,GoldenPRE2010,GoldenPRB2008, GoldenPRB2008E,Dillmann2012,HaghgooiePRE2005,JaiswalPRE2013} and references therein). The main purpose of this work is to put strong emphasis on the fact that the considered dipolar interaction belongs to the class of very soft long-ranged interactions, the limit opposite to the celebrated hard sphere interaction in three-dimensions (3D) and hard disc interaction in 2D. Based on this, a simple description of thermodynamic and dynamic properties is possible, using methods validated recently on other classical soft interacting particle systems, mainly in the plasma-related context.

Systems of soft interacting particles exhibit certain universal properties and there exist useful approximations, that are particularly suitable for this regime. In particular, the Rosenfeld-Tarrazona (RT) scaling~\cite{RT1,RT2} of the thermal component of the internal excess energy on approaching the freezing transition allowed previously to construct a very simple practical approach to the thermodynamics of weakly screened Yukawa systems in 3D~\cite{KhrapakPRE2015,KhrapakJCP2015,KhrapakPPCF2016}. An analog of the RT scaling also exists in the 2D case (although of a quite different functional form) and this has been recently used to construct a simple thermodynamic description of one-component plasmas and weakly screened Yukawa systems in 2D~\cite{KhrapakPoP2015,SemenovPoP2015,KhrapakCPP2016,KryuchkovJCP2017} with main applications to complex (dusty) plasmas. Here we apply the same arguments to the 2D system with $1/r^3$ dipolar interactions to put forward simple and accurate expressions for the thermodynamic properties of the liquid state, which are (by construction) in excellent agreement with the MD simulation results. Combined with the accurate calculation of the thermodynamic functions of the crystalline solid (using MD simulations and the shortest graphs method, proposed recently by some of the present authors) we are also able to approximately locate the fluid-solid phase transition, as well as the narrow coexistence region.

We also elaborate on the properties of collective modes in 2D dipolar systems. Recent investigations demonstrated that the quasi-crystalline approximation (QCA)~\cite{Hubbard1969,Takeno1971}, also referred to as the quasi-localized charge approximation (QLCA) in the plasma-related context~\cite{GoldenPoP2000}, is a good approximation to describe elastic collective modes in dense fluids for the regime of soft interactions (though it fails in the limit of very steep hard-sphere/hard-disc interactions~\cite{KhrapakSciRep2017}). Previously, QLCA has been applied with certain success to dipole-like systems in 2D~\cite{GoldenJPA2009,GoldenPRE2010}. Here we go somewhat further and derive simple analytic expressions, describing well the dispersion relations of the longitudinal and transverse elastic modes at sufficiently long wavelengths. The accuracy of these dispersion relations is demonstrated by comparing with the dispersions obtained from MD simulations. We demonstrate how these results can be useful in estimating the free energy of the crystalline solid. We also evaluate the high-frequency elastic moduli of the considered system and discuss relations to sound velocities operating in the strongly coupled fluid regime.

The rest of the article is organized as follows. In Section ~\ref{Methods} we  describe in detail the system under investigation, provide necessary details about the performed MD simulations, and summarize main thermodynamic relations used in this work. In section~\ref{Fluids} main results obtained for the fluid phase are summarized, including accurate expressions for thermodynamic quantities and detailed analysis of collective modes. In Section~\ref{Crystal} topics related to the thermodynamics of the crystalline phase are addressed. This includes thermodynamic functions, location of the fluid-solid phase transition, and sound velocities of an idealized lattice. Section~\ref{Moduli} is focused on elastic moduli and their relations to the sound velocities in a dense fluid phase. This is followed by our conclusion in Section~\ref{Concl}.

\section{Methods}\label{Methods}

\subsection{System description}
\label{SD}

We investigate a classical system of point-like particles in the 2D geometry, which are interacting via the pairwise repulsive inverse-third-power (IPL3) potential of the form
\begin{equation}\label{potential}
\phi(r) = \epsilon(\sigma/r)^3,
\end{equation}
where $\epsilon$ and $\sigma$ are the energy and length scales of the interaction. Phase behavior is conveniently described by the dimensionless interaction (coupling) parameter $\Gamma$,
\begin{equation}
\Gamma=\frac{\epsilon}{T}\left(\frac{\sigma}{a}\right)^3,
\end{equation}
where $T$ is the temperature (in energy units), $a=(\pi \rho)^{-1/2}$ is the 2D Wigner-Seitz radius, and  $\rho=N/V$ is the areal density of $N$ particles occupying the 2D volume (i.e. surface) $V$. The coupling parameter $\Gamma$ is roughly the ratio of the potential energy of interaction between two neighboring particles to their kinetic energy. The system is conventionally referred to as strongly coupled when the potential energy dominates, that is when $\Gamma\gg 1$.

At very low $\Gamma$ the system properties are similar to those of an ideal gas in 2D. When coupling increases the system forms a strongly coupled fluid phase, which can crystallize upon further increase in $\Gamma$. The nature of the fluid-solid phase transition in 2D systems depends considerably on the potential softness~\cite{KapferPRL2015}. For sufficiently steep repulsive interactions the hard-disk melting scenario holds: a first-order liquid-hexatic and a continuous hexatic-solid transition can be identified~\cite{BernardPRL2011,EngelPRE2013,ThorneyworkPRL2017}. For softer interactions the liquid-hexatic transition is continuous, with correlations consistent with the Berezinsky-Kosterlitz-Thouless-Halperin-Nelson-Young (BKTHNY) scenario~\cite{KapferPRL2015,KosterlitzRMP2017}. For the IPL family of potentials ($\propto r^{-n}$) the transition between the two regimes occurs at about $n\simeq 6$~\cite{KapferPRL2015}. The IPL3 system studied here thus belongs to the soft interaction class and the BKTHNY melting scenario should apply. This indeed has been observed both in numerical simulations~\cite{LinPRE2006} and colloidal experiments \cite{ZahnPRL1999,ZahnPRL2000,DeutschlanderPRL2014,KelleherPRE2015}.
However, the hexatic phase occupies a relatively narrow region on the phase diagram and its properties will not be addressed in this work.

\subsection{Computational details}
\label{MDdetails}

To obtain the accurate thermodynamic properties of IPL3 fluids and crystals in 2D, as well as necessary information about the properties of collective modes, extensive MD simulations have been performed using LAMMPS package~\cite{LAMMPS}. These MD simulations have been done in the $NVT$ ensemble at different temperatures using $N = 4 \times 10^4$ particles in a simulation box with periodic boundary conditions and the Nose-Hoover thermostat.  The initial systems configuration was chosen as an ideal hexagonal lattice and velocities were set according to the Maxwell distribution with the temperature equal to $1.5\,T$ and $2\,T$ for the fluid and crystal phases, respectively. The numerical time step of $\Delta t = 2.4 \times 10^{-4}\sqrt{ma^2\Gamma/\epsilon}$ was chosen. All simulation runs were performed for $6\times 10^5$ time steps, where the last $4\times 10^5$ steps were used for energy and pressure calculation based on standard functions implemented in the LAMMPS package (in the case of fluids, to guarantee that equilibrium was reached, we performed three simulations with different initial conditions for each examined state point). The cutoff radius of the potential was set equal to $25\rho^{-1/2}$. The internal energy and pressure were corrected accordingly, which resulted in a relative error of about $2\times 10^{-5}$, sufficient for the range of problems considered here.

\subsection{Thermodynamic relations}\label{Thermo}

The main thermodynamic quantities of interest in this work are the internal energy $U$, Helmholtz free energy $F$, and pressure $P$ of the system. The following thermodynamic definitions are useful~\cite{LL}
\begin{eqnarray}
U=-T^2\left(\frac{\partial}{\partial T}\frac{F}{T}\right)_V = \frac{\partial\left(F/T\right)}{\partial\left( 1/T\right)}|_V, \\
P=-\left(\frac{\partial F}{\partial V}\right)_T.
\end{eqnarray}
In addition, $U$ and $P$ can be calculated using the integral equations of state~\cite{HansenBook,Frenkel2001}
\begin{equation}\label{UPfromG}
\begin{split}
& U= N\left(T+ \frac{\rho}{2}\int{d\mathbf{r}\; \phi(r)g({\bf r})}\right),\\
& PV = N\left(T - \frac{\rho}{4}\int{d\mathbf{r}\; r\phi'(r)g({\bf r})} \right),
\end{split}
\end{equation}
where $g(\mathbf{r})$ denotes the radial distribution function, which is isotropic in gas and fluid phases and anisotropic in the crystalline phase.

We will use conventional reduced units: $u=U/NT$, $f=F/NT$, and $p=PV/NT$ and divide the thermodynamic quantities into the kinetic (ideal gas) and potential (excess) contributions, so that $u=1 + u_{\rm ex}$ (in 2D), $f=f_{\rm id}+f_{\rm ex}$, and $p=1+p_{\rm ex}$. Finally, it is useful to operate with the single coupling parameter $\Gamma$, instead of temperature and density. Since $\Gamma\propto a^{-3}T^{-1}\propto \rho^{3/2}T^{-1}$, the transformation of standard thermodynamic relations to their dimensionless form is governed by
\begin{equation}
\begin{split}
& \frac{\partial X}{\partial \rho}=\frac{3\Gamma}{2\rho}\frac{\partial X}{\partial \Gamma}, \\
& \frac{\partial X}{\partial T}=-\frac{\Gamma}{T}\frac{\partial X}{\partial \Gamma},
\end{split}
\end{equation}
where $X$ is a thermodynamic function of interest. In addition, a simple relation between the reduced excess pressure and energy for the IPL3 interaction in 2D holds:
\begin{equation}
p_{\rm ex}=\frac{3}{2}u_{\rm ex}.
\end{equation}
Other thermodynamic quantities can be readily evaluated when the excess internal energy is known. We summarize the main relations employed in this work in Appendix~\ref{A0}.

\section{Fluids} \label{Fluids}

\subsection{Thermodynamics of the fluid phase}\label{FluidsA}

The excess energy and pressure of the 2D IPL3 fluid have been determined using MD simulations. Based on these results, combined with our previous experience with thermodynamics of soft interacting particle systems in 2D geometry (mostly in the plasma-related context), simple and reliable analytical approximations are derived.

In the strongly coupled regime it is helpful to divide the thermodynamic properties, such as energy and pressure, into static and thermal contributions. The static contribution corresponds to the value of internal energy when the particles are frozen in a regular configuration and the thermal corrections arise due to deviations of the particles from these fixed positions (due to thermal motion). Here we relate the static energy to the lattice sum of the triangular lattice (Madelung energy) formed by particles interacting via the $\propto 1/r^{3}$ potential (this relation is meaningful for both crystalline and fluid phases). The corresponding lattice sum has been evaluated previously with a very high accuracy~\cite{Topping,derHoff}. The proportionality constant between the static energy and the coupling parameter (Madelung constant) is
\begin{displaymath}
M\simeq 0.798512.
\end{displaymath}
Thus, the excess internal energy in the fluid phase can be expressed as
\begin{equation}\label{f1}
u_{\rm fl} = M\Gamma + u_{\rm th}.
\end{equation}
The usefulness of this approximation stems from the fact that the ratio of the thermal-to-static contribution is small for strongly coupled  fluids with soft long-ranged interactions. In this case the static part is dominated by the cumulative contribution from large interparticle separations. This part is not very sensitive to the actual short-range order in the system since for large separations $g(r)\simeq 1$. It also does not change much across the fluid-solid phase transition, and thus the Madelung energy is an appropriate characteristic for both solid and fluid phases. Quantitatively, the static contribution is much larger than the kinetic energy ($M\Gamma\gg 1$), by the definition of strong coupling. On the other hand, the thermal contribution comes from the particle thermal motion and its magnitude is of the order of the average particle kinetic energy ($u_{\rm th}\sim 1$). This implies that even moderately accurate approximations for $u_{\rm th}$ result in a very accurate estimation of the total excess energy $u_{\rm fl}$ of strongly coupled fluids. The remaining step is therefore to identify an appropriate approximation for $u_{\rm th}$.

\begin{figure}[!t]
    \centering
    \includegraphics[width=85mm]{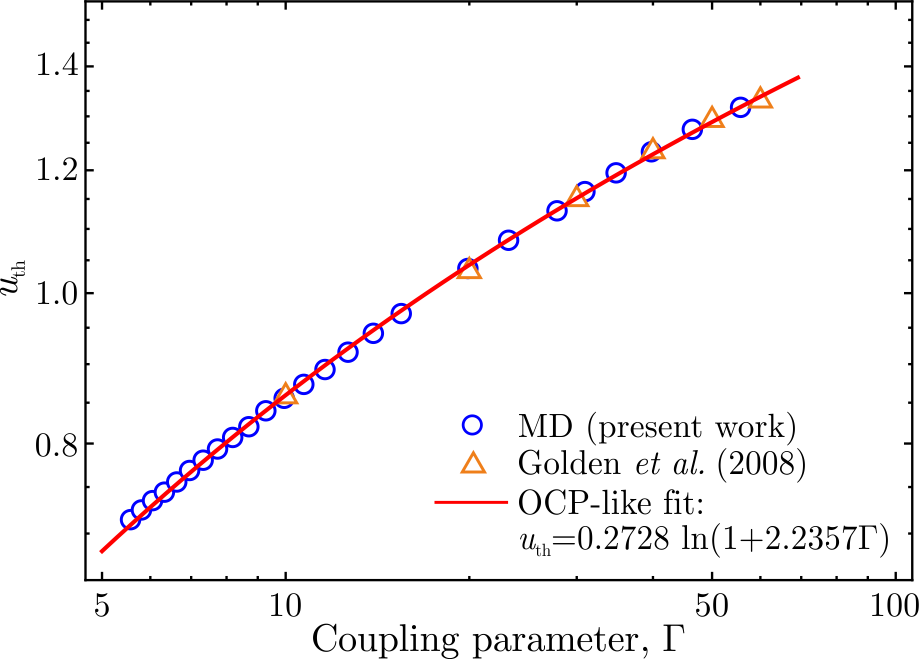}
    \caption{Thermal component of the reduced excess energy, $u_{\rm th}$, of a strongly coupled IPL3 system in 2D versus the coupling parameter $\Gamma$. Circles correspond to the results of MD simulations performed in this work. Triangles are the results by Golden {\it et al.}~\cite{GoldenPRB2008, GoldenPRB2008E}. The curve is the analytical fit of Eq.~\eqref{fit}.}
\label{IPL3-Figure1}
\end{figure}

Based on the previous results for other soft interacting particle systems in 2D (such as one-component-plasma~\cite{KhrapakPoP2014,KhrapakCPP2016} and weakly screened Yukawa systems~\cite{KhrapakPoP2015,KryuchkovJCP2017}), the thermal component of the excess energy is expected to exhibit a certain scaling with $\Gamma$ on approaching the fluid-solid transition. This scaling is to some extent analogous to the RT scaling of the thermal component of excess energy in 3D~\cite{RT1,RT2}, but has a quite different functional form. The functional form suggested for 2D systems with soft pairwise repulsive interactions is
\begin{equation} \label{fit}
u_{\rm th} =a \ln (1+b\Gamma).
\end{equation}
The validity of this functional form at sufficiently strong coupling is documented in Fig.~\ref{IPL3-Figure1}, where numerical data from the present MD simulations along with those reported previously~\cite{GoldenPRB2008,GoldenPRB2008E} are plotted. The best fit of the MD data obtained in this work yields $a=0.27284$ and $b=2.2357$. The fit is valid in the range $1\lesssim \Gamma \lesssim 70$. Combining Eqs.~(\ref{f1}) and (\ref{fit}) we write for the excess energy of the strongly coupled fluid phase
\begin{equation}\label{ufl}
u_{\rm fl}=M\Gamma+a \ln (1+b\Gamma).
\end{equation}
It is easy to ascertain that the first term is indeed dominant at strong coupling. For example, the ratio of the second to the first terms in Eq.~(\ref{ufl}) is $\simeq 0.4$ at $\Gamma=1$, it decreases to $\simeq 0.1$ at $\Gamma = 10$, and further drops to $\simeq 0.03$ at $\Gamma=50$. 

Having a fit for the reduced excess energy, we can evaluate the reduced excess Helmholz free energy from the relation
\begin{equation}\label{free_energy}
f_{\rm fl} = \int_0^{\Gamma}{d\Gamma'\; \frac{u_{\mathrm{fl}}(\kappa, \Gamma')}{\Gamma'}}.
\end{equation}
However, one must pay some attention to the fact that expression~(\ref{fit}) is not applicable all the way down to $\Gamma=0$. Although the actual contribution to the free energy from the weak coupling regime is of minor importance at strong coupling, we have accounted it in the following manner. At very weak coupling, the virial expansion allows us to estimate the free energy with a reasonable accuracy.
The first order correction to the ideal free energy is~\cite{LL}
\begin{equation}\label{SVC}
f_{1}=\frac{\rho}{2} \int\left[1-e^{-\phi(r)/T}\right]d{\bf r}=\int_0^{\infty}\left[1-e^{-\Gamma/x^3}\right]xdx,
\end{equation}
which can be evaluated analytically. 
%\begin{equation}\label{WeakCoupling}
%f_{1}=\frac{\pi \Gamma^{2/3}}{\sqrt{3}\Gamma(2/3)},
%\end{equation}
%where $\Gamma(x)$ in the denominator is the conventional gamma function (not to be confused with the coupling parameter $\Gamma$ appearing in the numerator). 
The excess energy in this regime is $u_{\rm ex}=\tfrac{2}{3}f_{1}$. However, careful comparison with the results from numerical simulations shows that Eq.~(\ref{SVC}) is applicable only at very weak coupling, $\Gamma\lesssim 0.05$. On the other hand, the strong coupling scaling (\ref{ufl}) is justified only for $\Gamma\gtrsim 1$. An approximation for the intermediate regime, basically based on an appropriate combination of Eqs.~(\ref{ufl}) and (\ref{SVC}) is described in Appendix~\ref{A}. Using this approximation [Eq.~(\ref{fit_lc})] below $\Gamma=10$ and Eq.~(\ref{ufl}) at higher values of $\Gamma$, and performing integration in Eq.~(\ref{free_energy}) we finally obtained a simple and accurate analytical approximation for the fluid free energy in the strong coupling regime, $\Gamma\gtrsim 10$:
\begin{equation}\label{ffl}
f_{\rm fl}=M\Gamma-a{\rm Li}_2(-b\Gamma)+0.381,
\end{equation}
where ${\rm Li}_2(x)$ is a polylogarithm function. It is the last constant term, which is responsible for the contribution from the weak-coupling regime. Clearly, the first two terms are dominant at strong coupling. 

Equations (\ref{ufl}) and (\ref{ffl}) represent our main results regarding thermodynamics of the IPL3 fluids in 2D. We will estimate below that the fluid-solid phase transition should occur at $\Gamma \simeq 69$, with a very narrow coexistence gap (see Section~\ref{Crystal}). Some thermodynamic quantities of the IPL3 melt (fluid just at the boundary of the fluid-solid coexistence) obtained using the approach discussed here are summarized in Appendix~\ref{A1}.

\subsection{Collective modes}\label{ColModes}

It is well known that fluids can exhibit different collective dynamics depending on the regime of coupling and correlations~\cite{HansenBook,TrachenkoRPP2016,FominJPCM2016,YangPRL2017}.
In the regime of weak correlations the dynamics is close to that in the ideal gas, and there exists only the longitudinal collective mode. On the other hand, in dense liquids not too far from the melting line, where interparticle correlations are strong, the transverse mode (one mode in the 2D case and two modes in the 3D case) can also be excited in addition to the longitudinal mode. It is this latter regime that will be mostly considered below.

A powerful theoretical approach to describe collective motion in classical systems of strongly interacting particles with soft pairwise interactions is the quasi-crystalline approximation~\cite{Hubbard1969,Takeno1971}. This approach can be considered  as either a generalization of the random phase approximation or, alternatively, as a generalization of the phonon theory of solids (the latter explains why it is often referred to as QCA). In the context of plasma physics an analog of the QCA is known as the quasi-localized charge approximation, QLCA. It was initially proposed as a formalism to describe collective mode dispersion in strongly coupled charged Coulomb liquids~\cite{GoldenPoP2000}. In recent years it was successively applied to strongly coupled one-component plasma in both 2D and 3D~\cite{GoldenPoP2000} as well as 2D and 3D Yukawa fluids, mostly in the context of complex (dusty) plasmas~\cite{RosenbergPRE1997,KalmanPRL2000,KalmanPRL2004,HartmannIEEE2007,
DonkoJPCM2008,HouPRE2009,UpadhyayaNJP2010,UpadEPL2011}. Application to the 2D IPL3 system was described in Refs.~\cite{GoldenPRE2010,GoldenJPA2009}. Here we discuss a procedure to derive simple explicit expressions for the longitudinal and transverse dispersion relations. We demonstrate that these dispersion relations are reasonably accurate at long wavelengths using the comparison with the results from MD simulations. Then, we also briefly discuss how the obtained results can be used to estimate the free energy of the IPL3 solid.

Within the QCA approach, the dispersion relations of elastic waves at strong coupling are directly expressed in terms of the radial distribution function (RDF), $g(r)$, and the pair interaction potential $\phi(r)$. The compact expressions are~\cite{Hubbard1969,Takeno1971,ZwanzigPRE1967}
\begin{equation}\label{w_L}
\omega_{\rm L}^2=\frac{\rho}{m}\int\frac{\partial^2 \phi(r)}{\partial z^2} g(r) \left[1-\cos(kz)\right]d{\bf r},
\end{equation}
\begin{equation}\label{w_T}
\omega_{\rm T}^2=\frac{\rho}{m}\int\frac{\partial^2 \phi(r)}{\partial y^2} g(r) \left[1-\cos(kz)\right]d{\bf r},
\end{equation}
where $\omega$ is the frequency, $k$ is the wave number, and $z=r\cos\theta$ is the direction of the propagation of the longitudinal mode (the particles are confined to the $zy$ plane). Here the subscripts ``${\rm L}$'' and ``${\rm T}$'' correspond to the longitudinal and transverse modes, respectively. The explicit expressions in the 2D case along with the expressions for the special case of the IPL3 system are provided in Appendix~\ref{B} for completeness. 

%The frequency spectra are thus determined by the first and second derivatives of the interaction potential $\phi(r)$ and the equilibrium fluid RDF $g(r)$; 

In the long-wavelength ($k\rightarrow 0$) regime the dispersion relations of the IPL3 fluid exhibit acoustic dispersion and the longitudinal and transverse sound velocities can be introduced,
\begin{equation}\label{Sound}
\lim_{k\rightarrow 0} \frac{\omega_{\rm L/T}^2}{k^2} = C_{\rm L/T}^2.
\end{equation}
Similarly to the IPL system in the 3D case~\cite{KhrapakSciRep2017}, these sound velocities can be easily related to the reduced excess energy (or pressure). For the considered case of IPL3 in 2D we immediately get for the QCA elastic sound velocities:
\begin{equation}\label{Acoust}
C_{\rm L}^2=\frac{33}{8}v_{\rm T}^2 u_{\rm ex}, \quad \quad C_{\rm T}^2=\frac{3}{8}v_{\rm T}^2 u_{\rm ex},
\end{equation}
where $v_T=\sqrt{T/m}$ is the thermal velocity of the particles.
Here we used the relation that follows directly from energy or pressure equations (\ref{UPfromG}),
\begin{displaymath}
p_{\rm ex}=\frac{3}{2}u_{\rm ex}=\frac{3}{4}\frac{\Omega_0^2a^2}{v_{\rm T}^2}\int_0^{\infty}\frac{g(x)dx}{x^2},
\end{displaymath}
where  $\Omega_0^2= 2\pi \rho\epsilon \sigma^3 /ma^3$ is the conventional 2D frequency scale.  Note an immediate consequence of Eq.~(\ref{Acoust}), $C_{\rm L}/C_{\rm T}=\sqrt{11}$, for the IPL3 system in 2D.

A simplest model $g(r)$, which takes into account the existence of a correlational hole (which prevent strongly repulsive particles from closely approaching each other) and is unity at longer separation (where correlations are small), turns out to be quite useful for soft repulsive interactions. Mathematically, this simplest model RDF reads
\begin{equation}\label{gOFr}
 g(x)=\theta(x-R),
\end{equation}
where $\theta(x)$ is the Heaviside step function and the radius of the correlational hole $R$ is of order unity in reduced units (the distances are expressed in units of $a$ here). A similar RDF was employed previously to analyze the main tendencies in the behavior of specific heat of liquids and dense gases at low temperatures~\cite{Stishov}. It was also used to calculate the dispersion relation of Coulomb bilayers and superlattices at strong coupling~\cite{Golden1993}. Physically, the model form of Eq.~(\ref{gOFr}) seems sensible in the present context, because the main contribution to the long-wavelength dispersion corresponds to long
length-scales, where $g(x)\simeq 1$. For soft enough interactions, this regime provides dominant contribution to the integrals in Eqs.~(\ref{IPL3_L}) and (\ref{IPL3_T}). The excluded volume effect for $x\leq R$ allows us to properly account for strong coupling. In addition, an appealing advantage of this simple RDF is that when substituted in the QCA (QLCA) expressions, it allows the analytical integration for certain interaction potentials. Particularly simple and elegant expressions have been recently derived for Yukawa systems and one-component plasma in 3D~\cite{KhrapakPoP2016,KhrapakJPA2017,KhrapakIEEE2017} and one-component plasma with logarithmic interactions in 2D~\cite{2DOCP}.

\begin{widetext}
For the considered IPL3 system in 2D the resulting expressions are not so elegant, although still tractable. We get for the longitudinal mode
\begin{equation}\label{R_L}
\omega_{\rm L}^2=\frac{\Omega_0^2}{R^3}\left\{ \frac{3}{2}-q^3R^3+\frac{J_1(qR)}{2qR}\left[6+2q^2R^2-2q^4R^4+\pi q^5 R^5 H_0(qR)
\right]
-\frac{J_0(qR)}{2}\left[
6-2q^2R^2-2q^4R^4+\pi q^4 R^4 H_1(qR)
\right]
\right\},
\end{equation}
where $q=ka$ is the reduced wave number, $J_0(x)$ and $J_1(x)$ are Bessel functions of the first kind, $H_0(x)$ and $H_1(x)$ are the Struve functions of order $0$ and $1$, respectively. The dispersion relation of the transverse mode is remarkably more simple,
\begin{equation}\label{R_T}
\omega_{\rm T}^2=\frac{\Omega_0^2}{R^3}\left\{
\frac{3}{2}-\frac{3J_1(qR)}{qR}
\right\}.
\end{equation}
\end{widetext}
The remaining step is the determination of the appropriate correlational hole radius $R$. Previously, it was proposed to determine $R$ from the condition that the model form (\ref{gOFr}) delivers good accuracy when substituted into the energy and/or pressure equations~\cite{KhrapakPoP2016}. This has been demonstrated to work well for both Yukawa and OCP systems in 3D and 2D OCP with logarithmic interactions~\cite{KhrapakPoP2016,2DOCP}. Following the same procedure we obtain
\begin{equation}
R=\frac{\Gamma}{u_{\rm ex}}.
\end{equation}
It is straightforward to verify that with this definition of $R$, the low-$q$ series expansion of Eqs.~(\ref{R_L}) and (\ref{R_T}) will reproduce the acoustic velocities given by (\ref{Acoust}). Note also that in the strong coupling regime the excess energy is mainly associated with the static contribution, $u_{\rm ex}\simeq M\Gamma$. In this regime the radius of the correlational hole is practically constant, $R\simeq 1/M\simeq 1.2523$.

\begin{figure*}[!t]
    \centering
   \includegraphics[width=135mm]{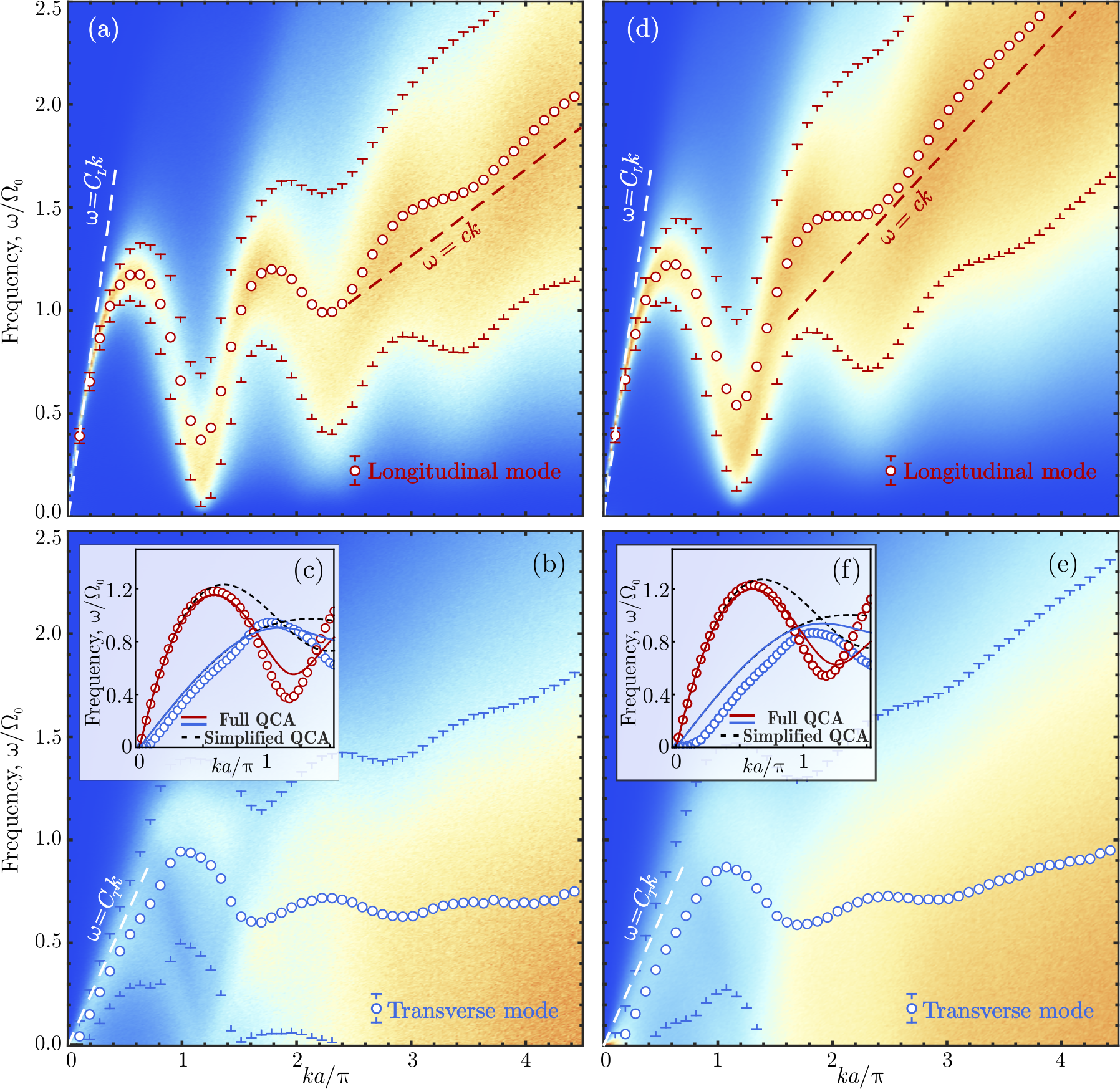}
    \caption{Longitudinal and transverse wave dispersion relations in color-coded format as obtained from Eq.~\eqref{wave_ampl} for $\Gamma=56$ (a) - (c)  and $\Gamma=28$ (d) - (f). Top panel [(a), (d)] corresponds to the longitudinal mode, bottom panel [(b), (e)] to the transverse mode. Circles, $\top$ and $\perp$ symbols correspond to the frequencies $\omega_0$ and $\omega_0\pm\alpha$ values, respectively, obtained by applying a fitting function \eqref{wave_ampl_fit} for every fixed value of dimensionless wave vector $q=ka$. White dashed lines correspond to the acoustic asymptotes $\omega=C_{\rm L} k$ and $\omega=C_{\rm T} k$, where $C_L$ and $C_{\rm T}$ are the  longitudinal and transverse sound velocities. For detailed discussion about the sound velocities in the IPL3 fluid, see Sec.~\ref{Moduli}. The dotted red lines correspond to the short-wavelength kinetic asymptote $\omega \simeq c k$, with $c=\sqrt{2}v_{\rm T}$ (see the text). The insets (c) and (f) show the long-wavelength portions of the dispersion relations. Here the red (blue) circles correspond to the longitudinal (transverse) dispersion relations as obtained from MD simulations. The red and blue solid curves display calculations using QCA approach of Eqs. (\ref{IPL3_L}) and (\ref{IPL3_T}) with the RDFs obtained from MD simulations. The dashed black curves correspond to simple analytical expressions of Eqs.~(\ref{R_L}) and (\ref{R_T}).}
\label{IPL3-Figure2}
\end{figure*}

In order to verify the quality of this simple analytical approximation, the dispersion relations of the IPL3 fluid have been obtained from MD simulations. We used the standard approach to compute phonon spectra in fluids, which is based on the longitudinal and transverse current correlation functions \cite{GoldenPRE2010,OhtaPRL2000,FominJPCM2016}.
Specifically, we calculated
\begin{equation}\label{wave_ampl}
	{\mathcal{A}}_{{\rm L},{\rm T}}(\mathbf{q},\omega) \propto q \mathrm{Re}\int dt\, \left<j_{{\rm L},{\rm T}}(\mathbf{q},t)j_{{\rm L},{\rm T}}(\mathbf{-q},0)\right> e^{i\omega t},
\end{equation}
where $j_{\rm L}(\mathbf{q},\omega)$ and $j_{\rm T}(\mathbf{q},\omega)$ are the projections of velocity current $\mathbf{j}(\mathbf{q},t)\propto\sum_i \mathbf{v}_i(t) \exp[i \mathbf{q} \mathbf{r}_i(t)]$ to longitudinal and transversal directions, respectively.
Here the summation is performed over all particles in the system,
$\mathbf{r}_i(t)$ is radius-vector of the $i$-th particle and $\mathbf{v}_i(t)$ is its velocity. Since fluids are isotropic we can average
${\mathcal{A}}_{{\rm L},{\rm T}}(\mathbf{q},\omega)$ over all directions of the wave vector $\mathbf{q}$ to get the dependence on $q=|\mathbf{q}|$.

Figure~\ref{IPL3-Figure2} shows, in color-coded format, the dispersion relations of the longitudinal and transverse waves obtained in this way for the two fluid state points characterized by $\Gamma=56$ (a) - (c) and $\Gamma=28$ (d)-(f). In contrast to crystals, color coding of current fluctuation spectra for fluids can merely  be used to illustrate qualitative properties. To get more quantitative information we fitted ${\mathcal{A}}_{{\rm L},{\rm T}}(q,\omega)$ by the Cauchy distribution,
\begin{equation}\label{wave_ampl_fit}
 f(\omega)\propto\frac{1}{(\omega-\omega_0)^2+\alpha^2}+\frac{1}{(\omega+\omega_0)^2+\alpha^2},
\end{equation}
for each value of $q$. Examples of the obtained dependencies $\omega_0(q)$ and $\alpha(q)$ are plotted in Fig.~\ref{IPL3-Figure2}. 

The long-wavelength portions of the dispersion relations $\omega_0(q)$ obtained from MD simulation are plotted in Figs.~\ref{IPL3-Figure2}(c) and \ref{IPL3-Figure2}(f). Here they are compared with QCA dispersion relations. The solid curves correspond to the ``full'' QCA with the actual RDF $g(r)$ obtained in MD simulations and substituted in Eqs.~\eqref{w_L} and \eqref{w_T}. The black dashed curves correspond to the ``simplified'' QCA given by analytical expressions (\ref{R_L}) and (\ref{R_T}). The agreement between the two versions of QCA and MD dispersions is satisfactory at sufficiently long wavelengths ($q\lesssim 2$ for the longitudinal and $q\lesssim 3$ for the transverse mode). This (low-$q$) regime corresponds to long length-scales, where both actual and model RDF are similar, $g(r)\simeq 1$.  For shorter wavelengths the two versions of QCA behave differently. This regime corresponds to short distances and the correct account of short-range correlations existing in liquids is necessary. Not surprisingly, the full QCA with the actual RDF is more consistent with MD-generated dispersion relations.  

Nevertheless, clear disagreement between the QCA and MD spectra is still observed at short wavelengths even with the use of accurate $g(r)$. The main reason for this is that QCA does not take into account effects of anharmonicity, which are causative, in particular, for damping of collective excitations. Indeed, the particles in liquid are considered within the framework of QCA as ``frozen'' near their equilibrium positions, whose statistics is determined by the actual RDF $g(r)$. Then, the excitation spectra are calculated in the harmonic approximation using perturbation theory for small displacements of particles around equilibrium positions. Account of particles' jumps (important for the physics of liquid state) cannot be done within the framework of perturbation theory~\cite{TrachenkoRPP2016}. At the same time, anharmonicity is related to the short-range region of the interaction potential, which corresponds to large $q$ in the reciprocal space and results in the observed growing discrepancy between the QCA and MD spectra. It should be also pointed out that the disappearance of the transverse mode at long wavelengths and the existence of a $q$-gap for the transverse waves propagation~\cite{YangPRL2017} cannot be properly described within the conventional QCA approach, because damping effects (associated with anharmonic interactions) are not included. One can see from examples presented in Figs.~\ref{IPL3-Figure2}(b) and (e) that the width of the $q$-gap decreases with increasing correlations (i.e. increasing $\Gamma$), in agreement with previous studies on collective excitations in various kinds of liquids~\cite{TrachenkoRPP2016,YangPRL2017,SchmidtPRE1997,BrykJCP2017}.    

Regarding the longitudinal mode, the asymptotic behavior of the dispersion relation at $q\gg 1$ is not described by QCA, because the kinetic effects are missing in this approximation. In this regime the characteristic scales of particle motion are much less than the average separation between the particles. The expected high-$q$ asymptote $\omega^2 \propto k^2 v_{\rm T}^2$ is consistent with MD simulation results. The proportionality coefficient $9/4$ has been previously suggested in Ref.~\cite{GoldenPRE2010}. Our preliminary analysis indicates that the coefficient $2$ can be more appropriate. The small ($\sim 10\%$) relative difference between these coefficients does not allow to discriminate between these two values using the obtained MD data. We, therefore, leave this issue for future work.   

\section{Crystals}\label{Crystal}

\begin{figure}[!t]
    \centering
    \includegraphics[width=85mm]{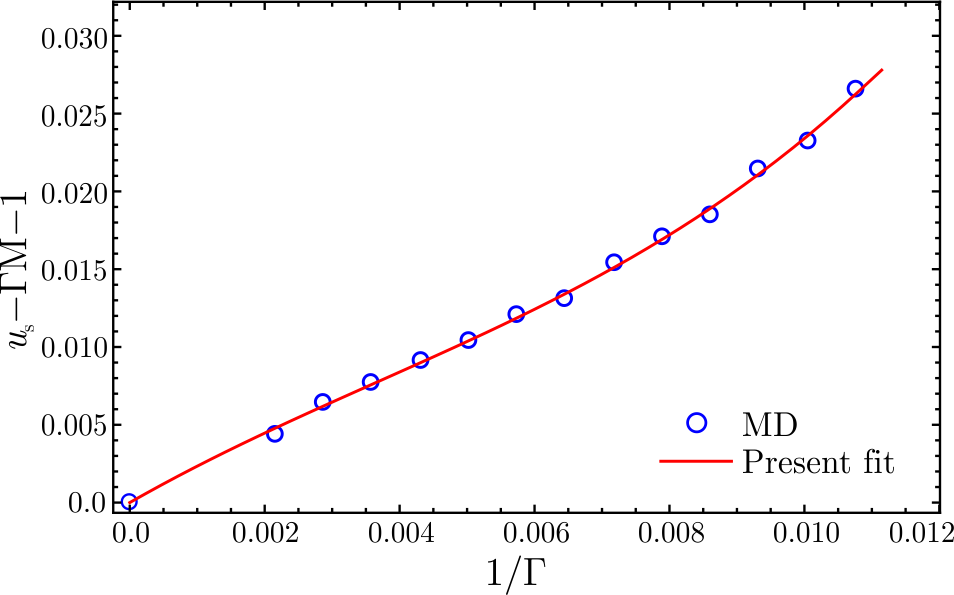}
    \caption{Anharmonic corrections to the reduced excess energy of the IPL3 crystal in 2D versus the inverse coupling parameter $1/\Gamma$.  Symbols represent the results from our MD simulation, the solid curve is a fit of Eq.~(\ref{crystal_fit}).}
\label{IPL3-Figure3}
\end{figure}

The reduced excess energy of a 2D crystalline lattice in the harmonic approximation is $u_{\rm h}=M\Gamma +1$.
We need to add a small anharmonic correction to get the total excess energy of a crystalline phase.
This anharmonic correction has been evaluated using MD simulations, and the results are plotted in Figure~\ref{IPL3-Figure3}. The anharmonic corrections are fitted using the standard functional form~\cite{DubinRMP1999}
\begin{equation}\label{crystal_fit}
u_{\rm anh}=\frac{A_1}{\Gamma}+\frac{A_2}{\Gamma^2}+\frac{A_3}{\Gamma^3}.
\end{equation}
The coefficients determined from the fitting are $A_1=2.47672$, $A_2=-148.77$, and $A_3=13507.4$ (curve in Fig.~\ref{IPL3-Figure3}). The resulting excess internal energy of the solid phase is
\begin{equation}
u_{\rm s}=M\Gamma+1+\frac{A_1}{\Gamma}+\frac{A_2}{\Gamma^2}+\frac{A_3}{\Gamma^3}.
\end{equation}

\begin{figure}[!t]
    \centering
    \includegraphics[width=85mm]{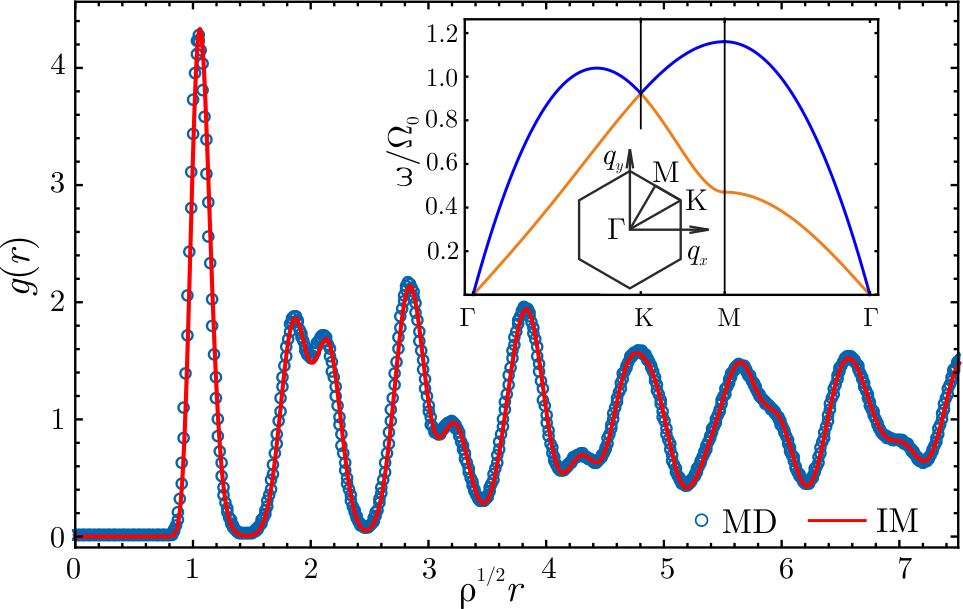}
    \caption{Example of the radial distribution function $g(r)$ for the IPL3 crystalline lattice at $\Gamma=92.8$. The symbols correspond to the MD simulation results, the solid curve is obtained using the IM approach. The inset demonstrates BvK phonon dispersion curves for the IPL3 triangular lattice.}
\label{IPL3-Figure4}
\end{figure}

The reduced excess Helmholtz free energy can then be evaluated in the following manner~\cite{HamaguchiPRE1997}. First, we divide it into anharmonic and harmonic contributions
\begin{equation}
 f_{\rm s} = \int\limits_{\infty}^{\Gamma}{d\Gamma'\; \frac{u_{\mathrm{anh}}(\kappa, \Gamma')}{\Gamma'}}+f_{\rm h},
\end{equation}
where $f_{\rm h}$ is the reduced excess free energy in the harmonic approximation. It is calculated by adding the lattice and vibrational free energies and subtracting the free energy of the perfect gas~\cite{LL,Alastuey1981}. In 2D geometry the resulting expression for the reduced harmonic free energy is~\cite{Alastuey1981}
\begin{equation}\label{fHarm}
f_{\rm h}= M\Gamma + \ln\Gamma + 1+ \frac{1}{2N}\sum_{{\bf k},{\rm s}}\ln\frac{\omega_{\rm s}^2(\bf k)}{\Omega_0^2},
\end{equation}
where $\omega_{\rm s}({\bf k})$ is the frequency of a phonon with wavenumber ${\bf k}$ and polarization ${\rm s}$, and the sum on ${\bf k}$ is over the first Brillouin zone in the reciprocal lattice. The sum of the last two terms is sometimes referred to as the harmonic entropy constant $\Sigma$, which is determined by the phonon spectrum of the crystalline lattice. The latter has been evaluated for the IPL3 triangular lattice using the standard technique, the resulting phonon dispersion curves are shown in the inset of Fig.~\ref{IPL3-Figure4}. The corresponding harmonic entropy constant has been evaluated as $\Sigma = 0.09284$ (a related approach to estimate $\Sigma$ using the QCA dispersion relations is briefly discussed in Appendix~\ref{C}). Thus, the reduced Helmholz free energy of the IPL3 crystalline solid is
\begin{equation}\label{f_sol}
f_{\rm s} = M\Gamma + \ln\Gamma+0.09284-\frac{A_1}{\Gamma}-\frac{A_2}{2\Gamma^2}-\frac{A_3}{3\Gamma^3}.
\end{equation}
This is our main result concerning the thermodynamics of the solid phase.

We can now estimate the location of the fluid-solid phase transition in the 2D IPL3 system by equating Helmholtz free energies of the corresponding phases. This yields $\Gamma_{\rm m}\simeq 69$ (here the subscript ``m'' refers to melting). In a more detailed consideration we equate fluid and solid temperatures, pressures and chemical potentials to evaluate the location and width of the phase coexistence region.
The standard procedure then yields $\Gamma_{\rm L}\simeq 69.2$ and $\Gamma_{\rm S}\simeq 69.4$. This is comparable to the results previously reported in the literature~\cite{ZahnPRL1999,ZahnPRL2000,LowenPRE1996,vanTeeffelenEPL2006,
JaiswalPRE2013} and is particularly close to the results from the Brownian dynamics simulations~\cite{HaghgooiePRE2005}. Note that a very narrow coexistence gap obtained, $\Delta\Gamma/\Gamma_{\rm S}\simeq 0.003$ (here $\Delta\Gamma = \Gamma_{\rm S}-\Gamma_{\rm L}$), should be related to the very soft character of the interaction potential. The anharmonic terms are not very important for the location of the phase transition: With neglecting anharmonic corrections, the fluid and solid free energy intersection point moves to $\Gamma_{\rm m}\simeq 75$. As a final remark, we note that we have not considered the existence of the hexatic phase. 

Thermodynamic properties of the IPL3 crystals can also be evaluated based on purely theoretical approach using an interpolation method (IM) for pair correlations in classical crystals proposed recently by some of the present authors~\cite{YurchenkoJCP2014, YurchenkoJCP2015, YurchenkoJPCM2016}. This approach allows us to compute RDF in the crystalline state based on the Born-von Karman (BvK) phonon spectrum and taking into account anharmonic corrections to the first correlation peak. The technical details of this approach are summarized in Appendix~\ref{IMethod}.

In Fig.~\ref{IPL3-Figure4} an example of the crystalline RDF $g(r)$ is presented. Visual similarity between the obtained theoretical RDF and MD data is the same as in previous applications of the IM approach~\cite{KryuchkovJCP2017,YurchenkoJCP2015,YurchenkoJPCM2016}. Using these highly accurate RDFs, pressure and excess energy can be readily obtained using Eqs.~\eqref{UPfromG}. In its simplest harmonic form ($\beta=0$) the IM approach provides a relative error in the excess energy smaller than $\simeq 10^{-3}$ in the worst case near the melting point. Taking into account anharmonicity, with the anharmonic correction coefficient $\beta=7.54$ [see Eq.~(\ref{Eq5})], obtained from MD simulations of the IPL3 crystal, reduces the relative error to $\simeq 10^{-4}$.

The longitudinal and transverse sound velocities of a perfect IPL3 2D lattice are given by Eqs.~(\ref{Acoust}) combined with $u_{\rm ex}=M\Gamma$ (we remind that QCA reduces to the standard phonon theory of solids in the limit $T=0$). The final result is
\begin{equation}\label{sound_crystal}
\begin{split}
& C_{\rm L}\simeq 1.8149 (\epsilon/m)^{1/2}(\sigma/a)^{3/2}, \\
& C_{\rm T}\simeq 0.5472 (\epsilon/m)^{1/2}(\sigma/a)^{3/2}.
\end{split}
\end{equation}

\section{High frequency elastic moduli and sound velocities in the liquid state}\label{Moduli}

The high frequency (instantaneous) elastic moduli for simple 3D fluids were derived by Zwanzig and Mountain~\cite{ZwanzigJCP1965}. The 2D analogues of these elastic moduli are
\begin{equation}
K_{\infty}=2\rho T-\frac{\pi\rho^2}{4}\int_0^{\infty}drr^2g(r)\left[\phi'(r)-r\phi''(r)\right],
\end{equation}
the high frequency limit of the bulk modulus~\cite{KhrapakOnset}, and
\begin{equation}
G_{\infty}=\rho T+\frac{\pi\rho^2}{8}\int_0^{\infty}drr^2g(r)\left[3\phi'(r)+r\phi''(r)\right],
\end{equation}
the high frequency limit of the shear modulus. The relations between the QCA elastic sound velocities derived in Section~\ref{ColModes} and the elastic moduli are:
\begin{equation}
m\rho C_{\rm L}^2= K_{\infty}+G_{\infty}-3\rho T, \quad\quad
m\rho C_{\rm T}^2= G_{\infty} -\rho T.
\end{equation}
Useful relations between $C_{\rm L}$ and $C_{\rm T}$ in both 3D and 2D have been recently discussed~\cite{QCA_Relations}. In particular, the high-frequency (instantaneous) sound velocity $C_{\infty}$, directly related to the instantaneous bulk modulus, can be introduced
\begin{equation}
C^2_{\infty}=K_{\infty}/m\rho = 2v_{\rm T}^2+C_{\rm L}^2-C_{\rm T}^2=v_{\rm T}^2\left(2+\frac{15}{4}u_{\rm ex}\right).
\end{equation}
The main purpose of this Section is to demonstrate that this instantaneous sound velocity is extremely close to the conventional adiabatic sound velocity appearing in hydrodynamic description of fluids~\cite{LL_Hydrodynamics}
\begin{equation}
C_{\rm s} =\frac{1}{m}\left(\frac{\partial P}{\partial \rho}\right)_{S} = v_{\rm T}\sqrt{\gamma \mu}.
\end{equation}

\begin{table}
\caption{\label{Tab1} Reduced sound velocities (in units of thermal velocity) of the IPL3 fluid in 2D evaluated for different values of the coupling parameter $\Gamma$, corresponding to the strongly coupled fluid phase.}
\begin{ruledtabular}
\begin{tabular}{llllllll}
$\Gamma$ & 10 & 20 & 30 & 40 & 50 & 60 & 70\\ \hline
$C_{\rm L}/v_{\rm T}$ & 6.04 & 8.38 & 10.18 & 11.70 & 13.04 & 14.25 & 15.37 \\
$C_{\infty}/v_{\rm T}$ & 5.93 & 8.11 & 9.81 & 11.24 & 12.51 & 13.66 & 14.72 \\
$C_{\rm s}/v_{\rm T}$ & 5.92 & 8.10 & 9.80 & 11.24 & 12.51 & 13.66 & 14.72 \\
\end{tabular}
\end{ruledtabular}
\end{table}

The sound velocities $C_{\rm L}$,  $C_{\infty}$, and $C_{\rm s}$ for several values of the coupling parameter $\Gamma$, corresponding to the strongly coupled fluid phase, are summarized in Table~\ref{Tab1}. It is observed that $C_{\rm L}$ overestimates the adiabatic sound velocity $C_{\rm s}$. However, the difference is rather small, as should be expected for the soft interaction potential studied in this work~\cite{KhrapakPPCF2016,KhrapakSciRep2017,KhrapakPRE2015_Sound}.
(For soft interactions at strong coupling one normally observe $C_{\rm L}\gg v_{\rm T}$ and $C_{\rm L}\gg C_{\rm T}$, which implies $C_{\rm L}\sim C_{\infty}\simeq C_{\rm s}$~\cite{KhrapakSciRep2017}). The instantaneous sound velocity $C_{\infty}$ is just slightly above the adiabatic sound velocity $C_{\rm s}$, the difference practically disappears with the increase in the coupling strength. The general inequality $C_{\rm s}\leq C_{\infty}$ was established by Schofield~\cite{Schofield1966}. We see that from the side of soft long-ranged interactions, this inequality is very close to equality, the observation previously reported for soft IPL systems in 3D~\cite{KhrapakSciRep2017}. This tendency, however, breaks down in the case of extremely steep hard-sphere-like interactions, where $C_{\rm L}$, and $C_{\infty}$ are all diverging (in 2D and 3D ~\cite{KhrapakSciRep2017,KhrapakJCP2016}), whilst $C_{\rm s}$ remains finite~\cite{RosenfeldJPCM1999}.

\section{Conclusion}\label{Concl}

We studied thermodynamics of two-dimensional IPL3 classical systems across coupling regimes, from the weakly non-ideal gas to the strongly coupled fluid and crystalline phases. Careful analysis of the extensive MD simulation results allowed us to put forward simple and physically suitable expressions for the thermodynamic properties (e.g. excess energy) of the investigated system. In particular, Helmholtz free energies of the fluid and solid phases have been derived and the location of the fluid-solid coexistence has been determined. The obtained results are comparable to those previously reported in the literature. A very narrow fluid-solid coexistence gap observed is likely related to the very soft nature of the interaction potential.

The QCA/QLCA approach has been applied to the description of collective modes of the IPL3 fluids. The use of a simplistic RDF has been suggested, based on previous results related  to strongly coupled plasma fluids. This has allowed us to derive explicit analytic dispersion relations for the longitudinal and transverse modes, which have been checked against the results of direct MD simulations. Reasonable agreement in the long-wavelength regime has been observed. We also briefly pointed out that the obtained simple fluid dispersion relations can be of some use in estimating the harmonic entropy constant of the solid phase.

The expressions for various sound velocities have been examined. These include conventional longitudinal and transverse elastic sound velocities of the idealized IPL3 crystalline lattice, their analogues (based on a QCA/QLCA approximation) in the strongly coupled fluid state, as well as conventional adiabatic sound velocity of the IPL3 fluid. Additionally, expressions for the 2D high frequency (instantaneous) elastic moduli have been introduced and related to the sound velocities. One useful observation is that the instantaneous sound velocity (related to the instantaneous bulk modulus) is extremely close to the adiabatic sound velocity. This observation is very likely a general property of soft long-ranged potentials, related neither to the exact shape of the interaction potential, nor to the dimensionality.

Finally, we would like to point once more that the interaction potential studied in this work represents just one particular example of very soft long-ranged interactions. It is therefore important that  the approaches used here can be directly (or with minor modifications) transferred and applied to other related soft-interacting particle systems.

\acknowledgments
This work has been supported by the A*MIDEX project (Nr.~ANR-11-IDEX-0001-02) funded by the French Government ``Investissements d'Avenir'' program managed by the French National Research Agency (ANR).
Numerical simulations and analysis have been supported by the Russian Science Foundation (RSF), Project No. 17-19-01691. Studies of fluid-solid transitions in 2D dusty plasmas and related model systems have been supported by RSF project No. 14-50-00124. We thank I. Semenov for careful reading of the manuscript.

\appendix

\section{Thermodynamic relations used in this work}\label{A0}

Here we express some of the reduced thermodynamic quantities of interest in terms of the reduced excess internal energy $u_{\rm ex}$. For example, the compressibility $Z=PV/NT$ is
\begin{equation}
Z=1+p_{\rm ex}=1+\frac{3}{2}u_{\rm ex}.
\end{equation}
The inverse reduced isothermal compressibility modulus $\mu=(1/T)(\partial P/\partial \rho)_T$ is
\begin{equation}
\mu = 1+\frac{3}{2}u_{\rm ex}+\frac{9\Gamma}{4}\frac{\partial u_{\rm ex}}{\partial \Gamma}.
\end{equation}
The reduced isochoric heat capacity $c_{\rm V}=(1/N)(\partial U/\partial T)_V$ is
\begin{equation}
c_{\rm V}=1+u_{\rm ex}-\Gamma \frac{\partial u_{\rm ex}}{\partial \Gamma}.
\end{equation}
The adiabatic index $\gamma = c_{\rm P}/c_{\rm V}$ is (for the considered potential and dimensionality)
\begin{equation}
\gamma = 1+\frac{(3c_{\rm V}-1)^2}{4\mu c_{\rm V}}.
\end{equation}
The quantities $\gamma$ and $\mu$ are used to calculate the conventional adiabatic sound velocity in IPL3 fluids. 

\section{Internal energy at intermediate coupling}\label{A}

\begin{figure}[!t]
    \centering
    \includegraphics[width=85mm]{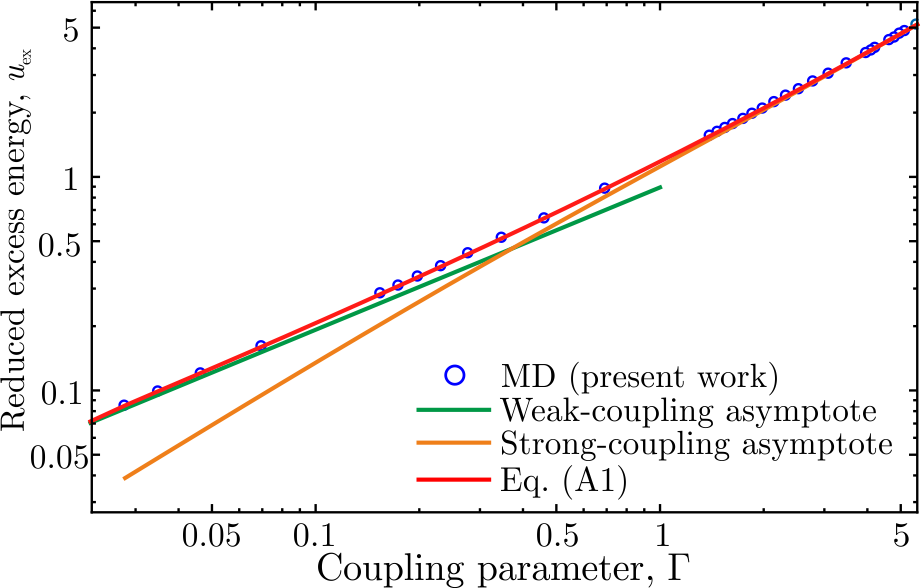}
    \caption{The reduced excess energy, $u_{\rm ex}$, of a weakly to moderately coupled IPL3 system in 2D versus the coupling parameter $\Gamma$. Circles correspond to the results of MD simulations performed in this work, the green line corresponds to the weak coupling asymptote, Eq.~(\ref{SVC}), the orange line corresponds to the strong coupling asymptote of Eq.~(\ref{ufl}), and the red curve is the fit of Eq.~(\ref{fit_lc}) appropriate for the moderately coupled regime.}
\label{IPL3-Figure5}
\end{figure}

The dependence of the reduced excess energy $u_{\mathrm{ex}}$ on the coupling parameter $\Gamma$ obtained from MD simulations at weak and moderate coupling is shown in Fig.~\ref{IPL3-Figure5} along with the scalings at strong coupling (\ref{ufl}) and weak coupling (\ref{SVC}). Based largely on the combination of these two scalings, we constructed a practical approximation (interpolation) suitable for the intermediate coupling regime. The proposed interpolation is
\begin{equation}\label{fit_lc}
u_{\rm ex} =\frac{2}{3} \left[1-\xi(\Gamma)\right] f_{1}+ \xi(\Gamma)\left[M\Gamma+A\ln(1+B \Gamma^s)\right],
\end{equation}
where
\begin{equation}
\xi(\Gamma)=\left[1+e^{-C(\Gamma-\Gamma_0)}\right]^{-1}.
\end{equation}
is the smooth step function.
Fitting MD data resulted in the following coefficients: $A=0.4791$, $B=1.2198$, $s=0.6044$, $C=428.216$, and $\Gamma_0=2.25\times 10^{-2}$. The corresponding curve is also plotted in Fig.~\ref{IPL3-Figure5} documenting excellent agreement with the MD results. 

\section{Thermodynamics quantities of the IPL3 melt}\label{A1}

In Table~\ref{Tab2} we have tabulated some reduced thermodynamic quantities of the IPL3 fluid near the boundary of the fluid-solid coexistence (IPL3 melt) at $\Gamma = 69$. The quantities displayed are internal thermal energy ($u_{\rm th}$), isochoric heat capacity ($c_{\rm V}$), adiabatic index ($\gamma= c_{\rm P}/c_{\rm V}$), Helmholtz free energy ($f_{\rm ex}$), excess internal energy ($u_{\rm ex}$), excess entropy ($s_{\rm ex}=u_{\rm ex}-f_{\rm ex}$), compressibility ($Z=1+p_{\rm ex}$), longitudinal elastic velocity ($C_{\rm L}/v_{\rm T}$), and transverse elastic velocity ($C_{\rm T}/v_{\rm T}$).

\begin{table}
\caption{\label{Tab2} Selected thermodynamic quantities (see the text for nomenclature) at the fluid boundary of the fluid-solid coexistence, at $\Gamma\simeq 69$.}
\begin{ruledtabular}
\begin{tabular}{lll}
$u_{\rm th}\simeq 1.38$ & $f_{\rm ex}\simeq 59.39$ & $Z\simeq 85.71$ \\
 $c_{\rm V}\simeq 2.11$ &  $u_{\rm ex}\simeq 56.47$ & $C_{\rm L}/v_{\rm T}\simeq 15.26$ \\
 $\gamma\simeq 1.02$ & $s_{\rm ex}\simeq -2.92$ & $C_{\rm T}/v_{\rm T}\simeq 4.60$   \\
\end{tabular}
\end{ruledtabular}
\end{table}

\section{Explicit expressions for the dispersion relations}\label{B}

Assume that a pairwise interaction potential can be written in the general form
\begin{displaymath}
\phi(r)=\epsilon_0 f(r/a),
\end{displaymath}
where $\epsilon_0$ is the energy scale [the subscript is used here to demonstrate difference in energy scales compared to Eq.~(\ref{potential}); for the IPL3 potential we have $\epsilon_0=\epsilon(\sigma/a)^3$]. The integrals
in Eqs.~ (\ref{w_L}) and (\ref{w_T}) can be simplified taking into account $d{\bf r}= rdr d\theta$ (in 2D), $kz=kr\cos \theta$ and that the derivatives of the interaction potential are
\begin{displaymath}
\frac{\partial^2 \phi}{\partial z^2}=\phi''(r)\frac{z^2}{r^2}+\frac{1}{r}\phi'(r)\left(1-\frac{z^2}{r^2}\right),
\end{displaymath}    
and
\begin{displaymath}
\frac{\partial^2 \phi}{\partial y^2}=\Delta \phi (r) - \frac{\partial^2 \phi}{\partial z^2}.
\end{displaymath}
Integration over the angle is then performed with the help of the identities
\begin{displaymath}
\frac{1}{2\pi}\int_0^{2\pi} \left[1-\cos(kr\cos\theta)\right]\cos^2\theta d \theta=\frac{1}{2}-\frac{J_1(kr)}{kr}+J_2(kr)
\end{displaymath}
and 
\begin{displaymath}
\frac{1}{2\pi}\int_0^{2\pi} \left[1-\cos(kr\cos\theta)\right]d \theta=1- J_0(kr),
\end{displaymath}
where $J_0(x)$, $J_1(x)$ and $J_2(x)$ denote the Bessel functions of the first kind, related via
\begin{displaymath}
J_0(x)+J_2(x)=\frac{2J_1(x)}{x}. 
\end{displaymath}
Introducing the reduced distance $x=r/a$ we obtain
\begin{widetext}
\begin{equation}\label{Gen_L}
\omega_{\rm L}^2= \Omega_0^2\int_0^{\infty} g(x)dx\left\{f'(x)\left[\frac{1}{2}-\frac{J_1(qx)}{qx}\right]+xf''(x)\left[\frac{1}{2}+\frac{J_1(qx)}{qx}-J_0(qx)\right]
\right\},
\end{equation}
and
\begin{equation}\label{Gen_T}
\omega_{\rm T}^2= \Omega_0^2\int_0^{\infty} g(x)dx\left\{f'(x)\left[\frac{1}{2}+\frac{J_1(qx)}{qx}-J_0(qx)\right]+xf''(x)\left[\frac{1}{2}-\frac{J_1(qx)}{qx}\right]\right\}.
\end{equation}
\end{widetext}
Here $\Omega_0^2=2\pi \rho\epsilon_0 /m= 2\pi \rho\epsilon \sigma^3 /ma^3$ is the nominal 2D frequency and  $q=ka$ is the reduced wave number. For the IPL3 potential with $f(x)=1/x^3$ the dispersion relations become   
\begin{equation}\label{IPL3_L}
\omega_{\rm L}^2= \frac{3\Omega_0^2}{2}\int_0^{\infty} \frac{g(x)dx}{x^4}\left[3-3J_0(qx)+5J_2(qx)\right],
\end{equation}
and
\begin{equation}\label{IPL3_T}
\omega_{\rm T}^2= \frac{3\Omega_0^2}{2}\int_0^{\infty} \frac{g(x)dx}{x^4}\left[3-3J_0(qx)-5J_2(qx)\right].
\end{equation}
To within some minor difference in notation, Eqs.~(\ref{IPL3_L}) and (\ref{IPL3_T}) coincide with Eqs. (16) and (17) from Ref.~\cite{GoldenPRE2010}. These expressions were previously used to generate the dispersion curves with the input of $g(r)$ data obtained in MD computer simulations~\cite{GoldenPRE2010,GoldenJPA2009}. A simplification, which does not require the accurate knowledge of the RDF, is discussed in Sec.~\ref{ColModes}.

\section{Harmonic entropy constant from QCA dispersion relations}\label{C}

Taking into account that we have two (longitudinal and transverse) modes in a 2D lattice and approximating the first Brillouin zone by a disk with the area $4\pi^2\rho$ we can re-write the harmonic entropy constant  [the last two terms in Eq.~(\ref{fHarm})] as
\begin{equation}\label{Harm1}
\Sigma=1+\frac{1}{4}\int_0^2\left[\ln\frac{\omega_{\rm L}^2(q)}{\Omega_0^2}
+\ln\frac{\omega_{\rm T}^2(q)}{\Omega_0^2}
\right]qdq,
\end{equation}
where $\omega_{{\rm L,T}}(q)$ correspond to the angularly averaged longitudinal and transverse phonon dispersion curves.
As a simplest rough estimate one can approximate the phonon spectrum by its isotropic acoustic asymptote, Eqs.~(\ref{Sound}) and (\ref{sound_crystal})  as was done in Ref.~\cite{Alastuey1981} for a 2D OCP with logarithmic interactions. In this way we have obtained for the present case of IPL3 in 2D
\begin{equation}\label{Sigma0}
\Sigma_0\simeq 0.6862,
\end{equation}
which significantly overestimates the actual harmonic entropy constant [subscript ``$0$'' in Eq.~(\ref{Sigma0}) denotes zero approximation]. In order to improve the accuracy we have also used the analytical QCA expressions (\ref{R_L}) and (\ref{R_T}) in Eq.~(\ref{Harm1}). This approach is based on the observation that angularly averaged lattice dispersion relations show remarkable similarity to isotropic QCA dispersion relation, in particular within the first Brillouin zone~\cite{Sullivan2006,HartmannIEEE2007}.
Taking $R=1/M=1.25233$ we have obtained in this approximation
\begin{equation}
\Sigma_{\rm QCA}\simeq 0.1336,
\end{equation}
which is considerably closer to the actual harmonic entropy constant. Thus, the fluid QCA dispersion relations in the strong coupling limit can be of some use in quickly estimating the free energy of the solid phase. Note that the harmonic entropy constant contributes only a small fraction of the total free energy for soft long-ranged interactions.

\section{Interpolation method for calculating RDF in 2D crystals}\label{IMethod}

The anisotropic RDF $g(\mathbf{r})$ of a crystal is written in the form \cite{YurchenkoJPCM2016}
\begin{equation}
\label{Eq1}
g(\mathbf{r}) = \frac{1}{\rho}\sum_\alpha{p_\alpha(\mathbf{r}-\mathbf{r_\alpha})},
\end{equation}
where the summation is over all the nodes $\alpha$, and
each individual peak has the shape
\begin{equation}
\label{Eq2}
\begin{split}
&p_\alpha(\mathbf{r}) \propto
 \exp\left[-\frac{\phi(\mathbf{r}+\mathbf{r_\alpha})}{k_{\rm B}T}-b_\alpha(\mathbf{e_\alpha}\cdot\mathbf{r})-
\right. \\
& \qquad\qquad \qquad \qquad \left.
-\frac{(\mathbf{e_\alpha}\cdot\mathbf{r})^2}{2 a_{\|\alpha}^2}-
\frac{\mathbf{r}^2-(\mathbf{e_\alpha}\cdot\mathbf{r})^2}{2 a_{\perp\alpha}^2}\right].
\end{split}
\end{equation}
The normalization constant as well as the parameters $a_{\|,\perp\alpha}^2, b_\alpha$ are defined by the conditions \cite{YurchenkoJPCM2016}
\begin{equation}
\label{Eq3}
\begin{split}
& \int{d\mathbf{r}\;p_\alpha(\mathbf{r})}=1, \qquad \int{d\mathbf{r}\;\mathbf{r}p_\alpha(\mathbf{r})}=0, \\
& \int{d\mathbf{r}\;(\mathbf{e_\alpha}\cdot\mathbf{r})^2 p_\alpha(\mathbf{r})}=\sigma_{\|\alpha}^2,\\
& \int{d\mathbf{r}\;[\mathbf{r}^2-(\mathbf{e_\alpha}\cdot\mathbf{r})^2] p_\alpha(\mathbf{r})}= \sigma_{\perp\alpha}^2,
\end{split}
\end{equation}
where $\mathbf{e_\alpha}=\mathbf{r_\alpha}/r_\alpha$ is the unit vector in the direction of $\mathbf{r_\alpha}$,
$\sigma_{\|,\perp}^2$ is the mean squared displacement for longitudinal and transverse directions, respectively.

The effect of the temperature dependence of phonon spectra can be taken into account by introduction of the anharmonic correction coefficient $\beta$~\cite{KryuchkovJCP2017}
\begin{equation}
\label{Eq5}
\sigma_{\|,\perp\alpha}^2 = \widetilde{\sigma}_{\|,\perp\alpha}^2 \left[1+\beta N\widetilde{\sigma}_{1}^2/V\right],
\end{equation}
where the tildes denote the mean-squared displacement (MSD) calculated using BvK phonon spectra (see Ref.~\cite{YurchenkoJCP2015}),
$\widetilde{\sigma}_1^2$ is the total MSD for the nearest neighbors.

Contrary to 3D crystals, in 2D  cases the mean squared displacements diverge logarithmically. The resulting correlation peaks become less localized, so the overlap of the neighboring peaks is generally stronger for 2D crystals. Nevertheless, it turns out that the
IM approach can be applied also in this case, in essentially the same way as for 3D crystals~\cite{YurchenkoJPCM2016}.

\bibliography{Ref-IPL3}

\end{document}